\documentclass[10pt,a4paper,oneside,onecolumn]{article}

\usepackage[caption=false,font=normalsize,labelfont=sf,textfont=sf]{subfig}

\usepackage{graphicx}
\usepackage{epstopdf}
\usepackage{cite}
\usepackage[cmex10]{amsmath}
\usepackage{multirow}
\usepackage{cleveref}
\usepackage{balance}
\usepackage{footnote}
\makesavenoteenv{tabular}
\usepackage{array}
\usepackage{dblfloatfix}

\usepackage{xcolor}

\date{}
\usepackage[left=2.30cm, right=2.30cm, top=2.30cm, bottom=2.30cm]{geometry}

\begin{document}

\title{Millimeter-wave Multimode Circular Array\\for Spatially Encoded Beamforming\\in a Wide Coverage Area}

\author{
	Stylianos D. Assimonis, M. Ali Babar Abbasi and	Vincent F. Fusco
	%	\\
	%	\thanks{
	\\
	{\normalsize
		Centre for Wireless Innovation (CWI),  Information
	}
	\\
	{\normalsize
	Institute of Electronics, Communications and Technology (ECIT)
	}
	\\
	{\normalsize
	Queen's University Belfast, University Rd, Belfast BT7 1NN, UK,
	}
	\\ 
	{\normalsize
		e-mail: \{s.assimonis,~m.abbasi,~v.fusco\}@qub.ac.uk
	}
}

\maketitle

\begin{abstract}
	
	This paper summarizes an investigation around millimeter-wave (mmWave) multimode circular antenna array based beamformer capable of specially encoded data transmission in a wide coverage area. 
	The circular antenna array is capable of an entire 360$^\circ$ azimuth sector coverage where broadcast, uni-cast and multi-cast radio transmissions are possible. 
	Orbital angular momentum (OAM) mode transmission along the elevation axis, i.e., perpendicular to the azimuth plane, can also be achieved. 
	These two types of spatially encoded beamforming makes coverage in almost an entire hemisphere possible. 
	The circular array is developed using twelve multilayer patch antennas with unique radiation performance aiding in both azimuth and elevation radiation. 
	The proposed array architecture and its radiation performance makes it a good candidate for spatially encoded beamforming for mmWave 5G.  
\end{abstract}

%\begin{IEEEkeywords}

\section*{Introduction and Motivation}

The development of modern communication systems led to the design, among others, of high-directive \cite{AssimonisR1}, compact \cite{AssimonisR2} antennas and beamformers \cite{AssimonisR3,AssimonisR4}. 
The latter, usually   require multiple radiating antennas whose excitation phase and magnitude variation can direct electromagnetic waves in the desired direction, forming radio beams. 
Uniform linear array (ULA) and uniform rectangular arrays (URA) are widely used in the first deployment phase of fifth generation of wireless communication (5G) radio access points, thanks to well known \textit{multiple--in--multiple--out} (MIMO) antenna technique \cite{SHAFI1,AssimonisR5}. Due to congestion in the radio frequency spectrum, the next deployment phase of 5G is now looking for best communication solutions at the millimeter-wave (mmWave) spectrum \cite{HONG1,book1}. The ULA and URA are limited in providing coverage along a maximum 180$^\circ$ sector area, while relatively less explored circular array geometry has a promising feature of providing coverage in entire 360$^\circ$ azimuth, making it a good choice for radio transmitters where 360$^\circ$ sectoral coverage is preferred. The claimed prospects of mmWave spectrum that are opening
new frontier in wireless communications with a promise for
the maximizing communication throughput using new antenna technologies motivate the investigation in this paper.

In this work, we focus on the wireless communication aspects of classical circular array theory \cite{Sheleg1968Matrix} and our follow-up investigations \cite{circ,oam1}. We first study how an array's antenna unit--cell operating in mmWave 5G band can be strategically designed in a way that when it is placed in a circular array formation, it can provide coverage not only along azimuth, but also along the elevation direction. We then focus on the conditions under which the electromagnetic waves with specific spatial encoding characteristics can be generated in mmWave spectrum. We continue our investigation by exploring broadcast, uni--cast, multi--cast, and OAM type spatially encoded radio transmissions.

\begin{figure}[t!]
	\centering
	\includegraphics[width=0.6\linewidth]{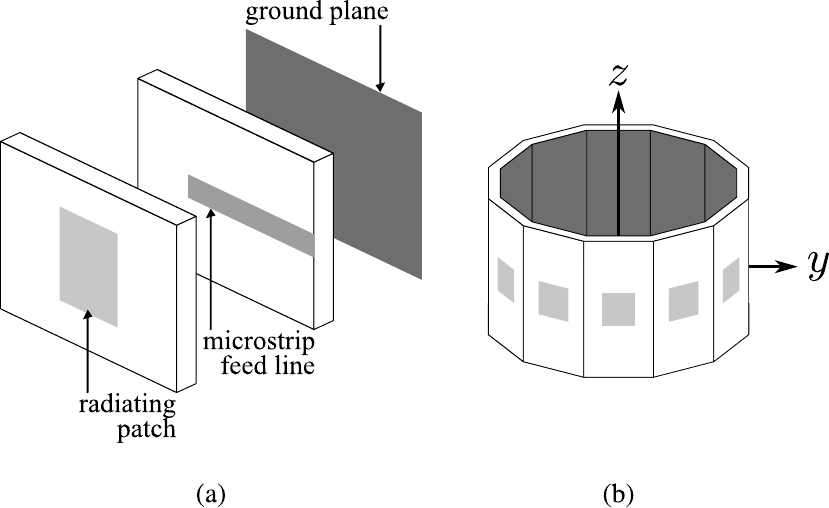}\vspace{-7pt}
	\caption{(a) Structural configuration of the microstrip patch antenna. (b) 12-element circular patch antenna array illustration.}\vspace{-10pt}
	\label{fig1}
\end{figure}

\section*{Multimode Circular Array}

The multimode circular array in this work comprises of $ 12 $ patch antenna elements. 
The latter configuration is shown in Fig. 1(a), developed using electromagnetically coupled patch, which is built using two substrate layers. 
The bottom substrate layer of the element is Rogers RO4003 having dielectric constant ($\epsilon_r$) of $ 3.38 $, and loss tangent ($\tan\delta$) of $ 0.0027 $, while the top substrate layer is Taconic TLY-5A with $\epsilon_r$ and $\tan\delta$ of $ 2.17 $ and $ 0.0009 $, respectively. 
A microstrip feed line sandwiched between TLY-5A and RO4003 substrates feed a radiating patch. 
The antenna is backed by a full ground plane as shown in Fig. \ref{fig1}a while it is terminated to a $ 50 $ $\Omega$ lumped port. 
Twelve copies of the same patch antenna are placed vertically to form a circular array, as illustrated in Fig. \ref{fig1}b. 

\begin{figure}[t!]
	\centering
	\includegraphics[width=0.7\linewidth]{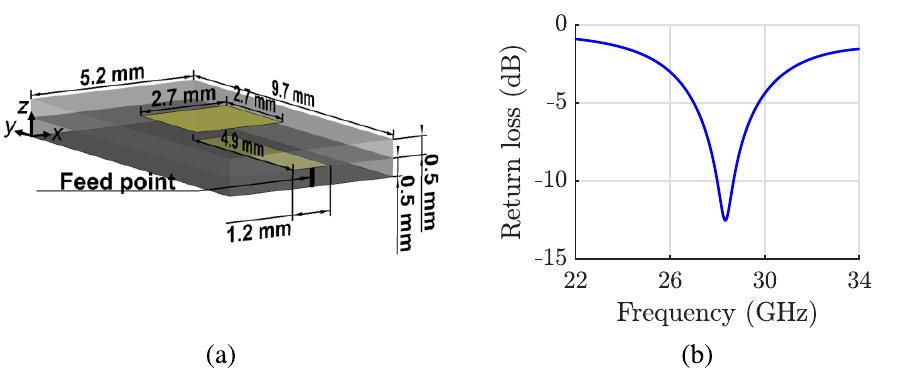}\vspace{-7pt}
	\caption{(a) Design parameters of the multi-layer microstrip patch antenna element, and (b) return loss at the feed point.}
	\label{fig2}
\end{figure}

The antenna element is designed to operate at $ 28 $ GHz through full electromagnetic analysis in CST STUDIO SUITE.
The obtained element's dimensions are presented in Fig. \ref{fig2}a, while its simulated impedance matching performance at $ 50 $ $ \Omega $ is depicted in Fig. \ref{fig2}b:  please note that  we show only the return loss of a single element due to the symmetrical geometry of the circular array. 
Simulated radiation efficiency of the antenna element is $98\%$.
Note that there is a gap between radiating patch and the feed line, which offers a unique radiation feature. 
It not only ensures radiation along $+z$-direction, as expected from patch antennas \cite{wong2004compact}, but also along $+y$-direction (relative to the Cartesian coordinates shown in Fig. \ref{fig2}a). 

\begin{figure}[t!]
	\centering
	\includegraphics[width=0.6\linewidth]{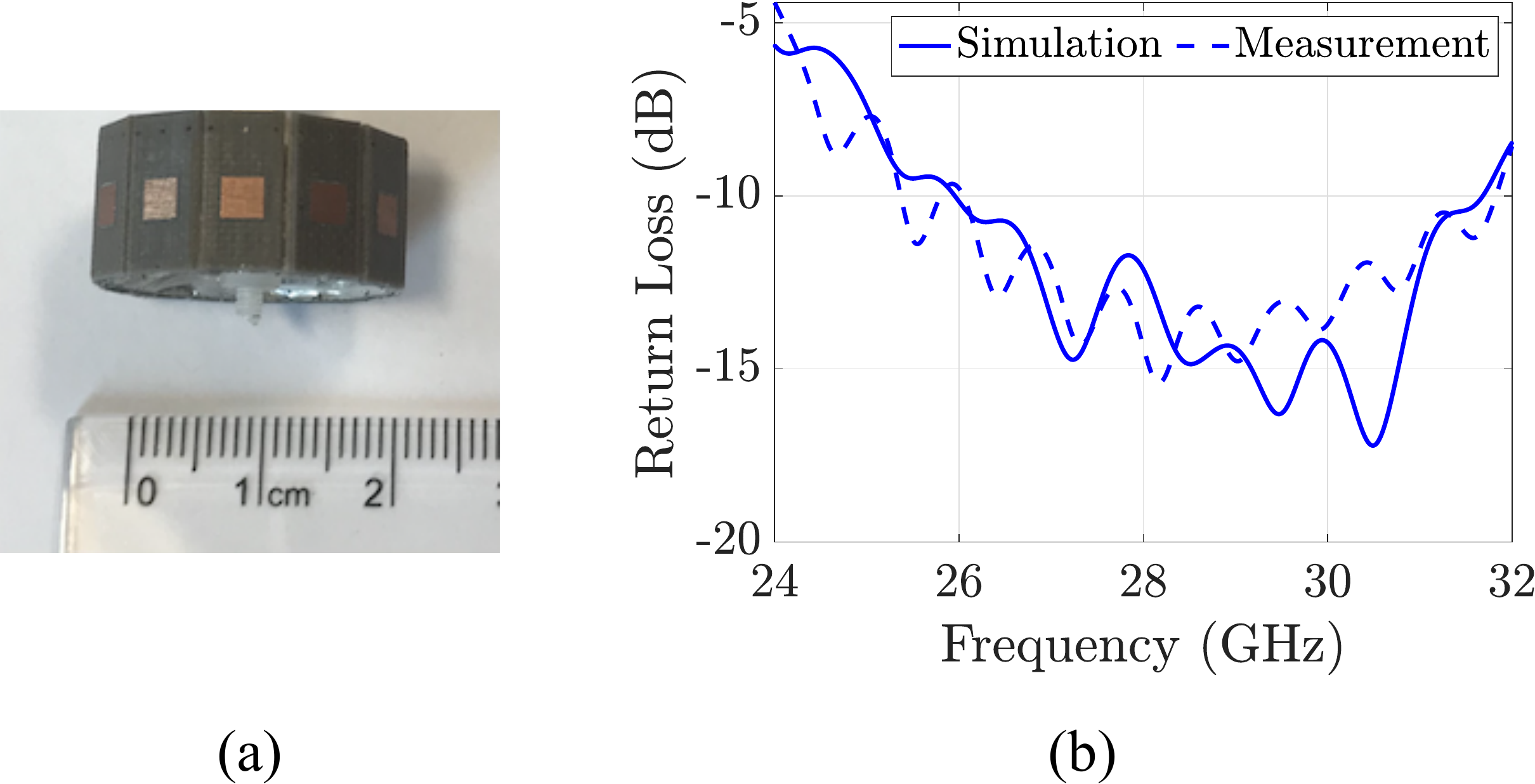}\vspace{-7pt}
	\caption{(a) Fabricated circular antenna array and (b) return loss at the feed point of a single element, when the other $ 11 $ are open-circuited.}
	\label{fig2_2}

\end{figure}

Next, $ 12 $ of the these antenna elements are utilized in the development of the circular array, which has diameter of $ 19.38 $ mm (i.e., $1.8\lambda$), and placed in $xy$-plane to operate as a radio transmitter. 
Photograph of the fabricated and measured prototype is also shown in Fig. \ref{fig2_2}a. 
Fig. \ref{fig2_2}b depicts the return loss of a single element of the antenna when all the others are open circuited: a good agreement between simulated and measured results is observed. 
Due to strategic designing of the antenna elements, as previously mentioned, it will not only provide coverage along azimuth direction $xy$-plane, but also along the elevation direction, i.e. towards $+z$-direction (Fig. \ref{fig1}b), covering a wider area compared to a circular array designed using standard patch antennas \cite{wong2004compact,oam2}, as it will be explained.

\section*{Spatially Encoded Beamforming}

\begin{figure}[t!]
	\centering
	\includegraphics[width=0.8\linewidth]{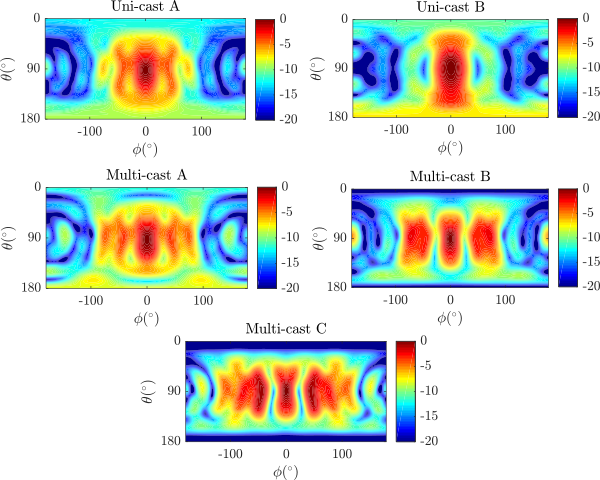}
	\caption{Simulated normalized directivity patterns for multimode excitation of the circular array resulting in uni- and multi-cast transmission.}
	\label{fig3}
\end{figure}

In this section, we study how radiated signals from the circular array in Fig. \ref{fig1} can be spatially encoded. First along the azimuth direction, it has been thoroughly discussed in \cite{circ} that spatially encoded signal transmission can be achieved via mode-mixing excitation of circular array elements. Following the same theoretical background, we show that the mode-mixing excitation of circular array allows broadcast, uni-cast and multi-cast radio transmissions along the azimuth direction. In Fig. \ref{fig3}, multiple cases of uni-cast and multi-cast directivity patterns are shown, that are results of mode-mixing excitations from the subsets of $m = 0, \pm1, ...\pm5$, where $m$ represents the $m$-th mode excited at the circular array. 
Please refer to \cite{circ,Sheleg1968Matrix} for the derivation of the $m$ mode--mixing excitations. Uni-cast radiation has two cases, A and B, in both of them, high directivity beam can be observed along $\phi = 0^\circ$, which can be used to serve single user. 
Simulated directivity beam maxima for the two uni-cast cases are $ 8.59 $ and $ 8.91  $dBi, that is higher compared to broadcast radiation, where maximum directivity usually takes values around $ 4 $ dBi \cite{circ2}. 
To serve multiple users simultaneously, three multi-cast cases (A, B and C) are shown in Fig. 3 with three high directivity beams along three dominant directions. Simulated peak directivity for the multi--cast beams are in the range of 6.5 to 7.6 dBi. Note that the uni--cast and multi--cast beams are rotatable along entire 360$^\circ$ azimuth directions, thanks to the array's circular symmetry.

\begin{figure}[t!]
	\centering
	\includegraphics[width=0.4\linewidth]{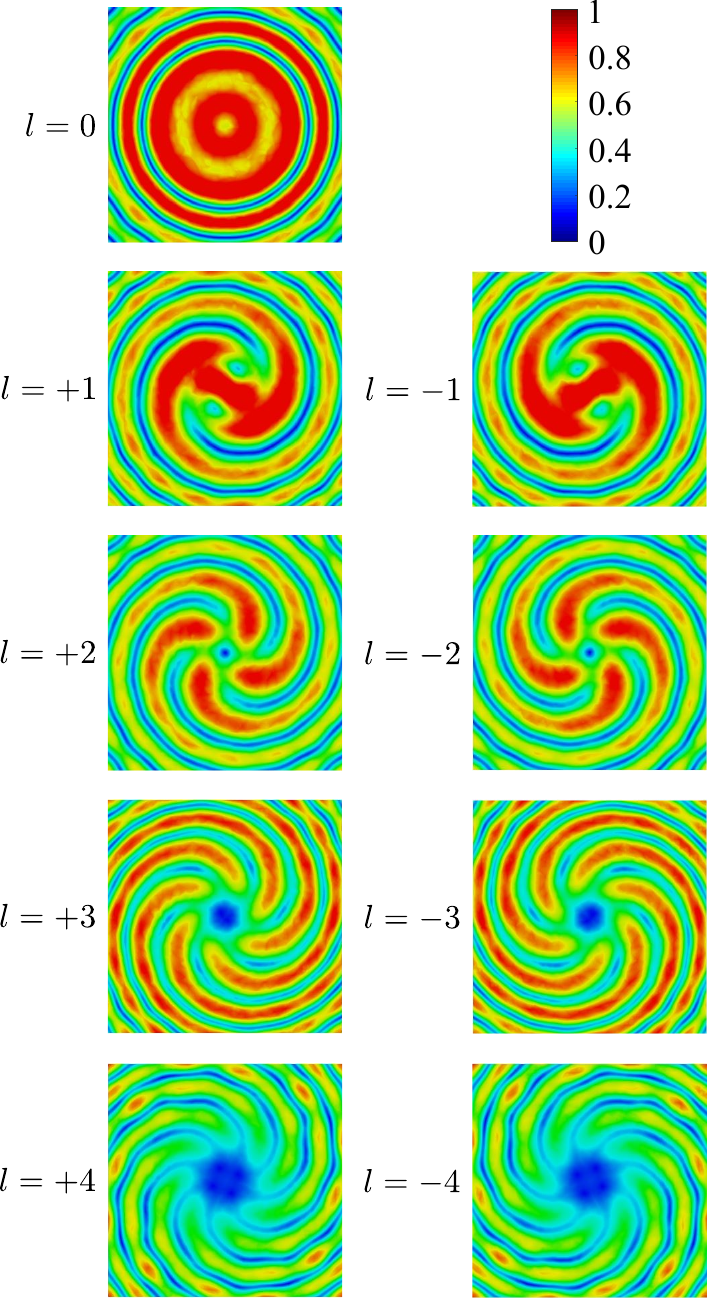}\vspace{-5pt}
	\caption{Snap shots of the normalized $E$-field captured on a plane 2$\lambda$ above the circular array representing spiral wave propagation when $\ell$ modes are excited.}
	\label{fig4}
\end{figure}

Antenna elements are placed in circular array formation builds a series of radiating slots tangential to the array circle when observed from array's top side. Excitation of these slot formations result in cavity modes, aiding in radiation along the $+z$-direction. 
We observe that when OAM modes are excited, we can realize classical vortex beam type radiation along the $+z$-direction. Mode excitation of OAM is generally referred to as $\ell$, whose mathematical formulation can be found in \cite{oam2,oam3}. 
Excitation of all antennas using same magnitude and phase signal results in the first OAM mode ($\ell = 0$) along $+z$-direction, which is analogous to the broadcast radiation along azimuth direction ($m = 0$). 
Increasing the excitation to $\ell = \pm1, \pm2, \pm3$ so on results in the formation of vortex beams with vortex along $+z$-direction and spherical OAM phase ramps \cite{oam3}. The cycles of phase spiral ($0^\circ - 360^\circ$) depends upon the excitation mode $\ell$, while the magnitude and phase radiation gives a rise to spatially encoded data transmission. 

To observe the radiation performance in elevation against OAM mode excitation, we simulated the $ E $-field in a plane parallel to azimuth plane, which lies above the circular array at distance of 2$\lambda$ .
This allow us to understand how 28 GHz electromagnetic energy radiates from the circular array when a certain $\ell$-th modes is excited. The resultant animated field files against $\ell$ mode excitations are available at the shared link \footnote{{\fontfamily{pcr}\selectfont{go.qub.ac.uk/mmWaveCircularArray}}}, while snapshots the animated fields at a radiated field angle of 0$^\circ$ is provided in Fig. \ref{fig4}. 

Although, for certain array configurations traditional MIMO theory leads to eigen-modes identical to OAM type radiation, as argued in \cite{oam2}, we used widely understood term ``OAM'' for the spatial encodings observed from the animated fields and Fig. \ref{fig4}. 
Although data transmission using separate OAM channels is realizable, the radiation mechanism can be generalized using MIMO theory since any antenna array geometry, including the one we used in this investigation, i.e., circular array, can be handled by a MIMO system. 
The main different is that in a MIMO case, often omni--directional radiators are considered, whereas here we develop circular antenna array using patch antenna excited by a feed line and backed by a full ground, which entirely changes the radiation mechanism when it is compared with a circular array build using ideal omni-directional radiators. 
Nevertheless, the sets of results presented in Fig. \ref{fig3} and \ref{fig4} indicate that the strategically designed circular array can be used for a variety of spatially encoded beam types governed by $m$ and $\ell$ excitations.

%\begin{figure}[]
%	
%	\centering
%	\subfigure[]{\centering\includegraphics[width=5cm]{../DataAndFigs/OAM_Figure1.pdf}}\hspace{7pt}
%	\caption{Snap shots of the $E$--field captured on a plane 2$\lambda$ above the circular array representing spiral wave propagation when $\ell$ modes are excited.}
%	\label{fig4}
%\end{figure} 

\section*{Conclusions and Future Objectives}

In this paper, we summarize a developing investigation on mmWave multimode circular array for spatially encoded beamforming in 360$^\circ$ azimuth direction and close-to-hemispherical coverage along elevation. In the azimuth direction, the circular array is capable of generating uni-- and multi--cast mmWave radio beams, while along the elevation direction, it is capable of generating multiple OAM vortex beams. The array is useful for spatially encoded data streams transmission in a wide coverage. Simple fabrication process make the proposed circular array structure a good candidate for 5G and beyond wireless applications. Our next objective is to test the radiation and mode-orthogonality performance of the array when it is a part of mmWave transmitter system.

\section*{Acknowledgment}

This work was supported by the EPSRC, U.K., under grant EP/P000673/1 and grant EP/EN02039/1. Authors would like to thank K. Rainey for assisting in the hardware development and measurements.

\end{document}